%% file: main.tex
\documentclass[sigconf,screen]{acmart}
\AtBeginDocument{%
  }

\setcopyright{cc}
\setcctype{by-nc-nd}
\acmDOI{10.1145/3805760.3814904}
\acmYear{2026}
\copyrightyear{2026}
\acmISBN{979-8-4007-2601-9/2026/07}
\acmConference[AIware '26]{Proceedings of the 3rd ACM International Conference on AI-Powered Software}{July 6--7, 2026}{Montreal, QC, Canada}
\acmBooktitle{Proceedings of the 3rd ACM International Conference on AI-Powered Software (AIware '26), July 6--7, 2026, Montreal, QC, Canada}
\acmSubmissionID{fseaiware26main-pp034-p}
\received{2026-02-15}
\received[accepted]{2026-03-28}

\usepackage{tcolorbox}
\usepackage{enumitem}
\usepackage{listings}
\usepackage[table]{xcolor}
\usepackage{tabularx}
\usepackage{xcolor}
\usepackage{float}
\usepackage{booktabs}
\usepackage{balance}

\begin{document}

\title{From Correctness to Consistency: Redefining Reliability for the Agentware Era}

\author{Xue Qin}
\affiliation{%
  \institution{Villanova University}
  \city{Villanova}
  \country{United States}}
\email{xue.qin@villanova.edu}

\author{Maurício Gruppi}
\affiliation{%
  \institution{Villanova University}
  \city{Villanova}
  \country{United States}}
\email{mgouveag@villanova.edu}

\renewcommand{\shortauthors}{Xue et al.}

\begin{abstract}
  The rise of ``vibe coding'' has marginalized requirement analysis, allowing unverified ``Make it Work'' assumptions to accumulate into complex defects that evade standard testing. In the absence of a reliable ground truth oracle, we advocate a paradigm shift from correctness verification to logic consistency. We introduce Differential Logic Analysis (DLA), a framework that utilizes Logical Inference to detect internal contradictions across parallel and sequential development workflows. Preliminary simulations demonstrate that DLA successfully intercepts logic drift, such as contradictions and unspecified assumptions, and offers a new perspective on reliability assessment in the AI-assistant development era.
\end{abstract}






\maketitle

\input{content/intro}

\input{content/method}

\input{content/results}

\input{content/discussion}
\input{content/related}

\input{content/conclusion}

\balance
\bibliographystyle{ACM-Reference-Format}
\bibliography{references}


\end{document}

%% file: content/intro.tex
\section{Introduction}
\label{sec:introduction}

The proliferation of Large Language Models (LLMs) like GitHub Copilot~\cite{copilot2026} and ChatGPT~\cite{chatgpt2026} is transforming software construction, enabling non-experts to build complex systems via natural language~\cite{openai2024gpt4technicalreport, Gerald2024IeeeAccess}. This has given rise to ``vibe coding,'' where developers orchestrate AI components through rapid iteration rather than rigorous planning~\cite{peng2023impactaideveloperproductivity}. However, this speed-driven approach has created a critical gap: the systematic abandonment of Requirement Analysis. Practitioners frequently bypass formal specifications, only to find that their initial instant success~\cite{TechTargetVibeCoding, NBC2025VibeCoding} often degrades into prolonged struggles with prompt revision and complex bugs. As a result, developers increasingly report abandoning projects entirely rather than attempting to debug AI-generated logic~\cite{fawzy2025vibecodingpracticemotivations}.

This volatility stems from the stochastic non-determinism of LLMs~\cite{song-etal-2025-good}. 
To bridge the gap between prompts and execution, models inject ``Make it Work'' assumptions that are often explicitly or implicitly flawed~\cite{wang2025Assumptions}. 
This gap arises because prompts provide high-level, often underspecified instructions, while the resulting implementations require implicit assumptions to become executable. 
Consequently, iterations compound hallucinations~\cite{zheng2025isstaLLMHallucinations} rather than refining them. 
Testing code generated by an LLM-assistant fails to break this cycle due to an inherent logical bias: when the same model generates both code and tests, the latter inherits the same unverified assumptions~\cite{ma2025rethinkingverificationllmcode}, creating a self-validating loop.

Moreover, the gap between prompt and execution exposes a deeper challenge of the Agentware era: the absence of a reliable ground truth.
In traditional software engineering, implementation correctness can be validated against formal specifications and detailed requirements. 
An AI coding assistant may make assumptions that do not reflect the developer’s intent and are not grounded in an explicit specification.
There is no authoritative reference against which the assumptions can be verified. As a result, the intended system behavior becomes ambiguous and potentially conflicted.
We therefore advocate a paradigm shift from correctness verification to \emph{logic consistency}.
Rather than asking if artifacts are ``correct'' relative to a missing specification, we examine whether their internal logical commitments remain coherent. 
To operationalize this, we introduce \textbf{Differential Logic Analysis (DLA)}, a logic inference framework \cite{wu2024survey} that treats prompts and code as comparable premises to detect logical divergence. 
By identifying contradictions and underspecified interactions, DLA intercepts drift across both parallel and sequential development scenarios.
To explore the viability of this framework, we address two core research questions:
\begin{itemize}[leftmargin=*, nosep]
    \item RQ1: Can DLA identify logical contradictions in LLM artifacts that bypass auto-generated test suites?
    \item RQ2: How does logic divergence manifest differently across various workflows?
\end{itemize}

To answer the questions, we conducted two case studies:
1) Parallel Development: Where multiple agents independently implement related features, creating the potential for logic divergence~\cite{HuthRyan2004} in 
isolation (e.g., assumptions made by agents work locally but conflict globally).
2) Sequential Iteration: Where a single agent refines a codebase over multiple turns, creating the potential for logic drift over iterations (e.g., new-turn logic diverges from prior logic).
We then manually applied DLA rules to the artifacts generated in these simulations. By treating the natural language prompts and code segments as logical premises, we tested whether Logical Inference,
specifically \textit{Entailment} (logical consequence), \textit{Contradiction}, and \textit{Neutrality}~\cite{MacCartney2008nli}, could successfully contextualize software logic. Our analysis confirmed that DLA could systematically identify ``logic divergence'' and ``unspecified assumptions'' that auto-generated tests failed to catch, proving that \textbf{Logic Consistency} is a viable and measurable proxy for reliability in the absence of a ground truth. 
In this vision paper, we contribute:
\begin{itemize}[leftmargin=*, nosep]
    \item \textbf{A New Perspective (Consistency $>$ Correctness):} We advocate shifting correctness-based verification with logic consistency checks to ensure coherence when formal oracles are unavailable.
    \item \textbf{The Differential Logic Analysis (DLA) Framework (RQ1):} We introduce a logic-inference methodology adapted from formal methods~\cite{Hoare1969} to detect conflicting assumptions in parallel and sequential AI-assisted development.
    \item \textbf{Feasibility Evidence (RQ2):} We provide empirical evidence that logic divergence is a distinct, detectable failure mode that persists even when code passes syntax and runtime checks.
\end{itemize}

%% file: content/method.tex
\section{Study Methodology}
\label{sec:methodology}

\subsection{Logic-Centric View of AI-Assisted Artifacts}
\label{sec:logic_view}

Traditional software engineering relies on rigorous specifications to bridge the gap between intent and code. 
In contrast, AI-assisted development often bypasses formal requirements, relying on Large Language Models (LLMs) to bridge sparse natural language prompts to executable instructions. 
This shift introduces a ``Logic-Consistency'' problem where the LLM must fill specification gaps with unverified assumptions.
To support this new development lifecycle, we propose a \textit{Differential Logic Analysis (DLA)} to audit the alignment among artifacts. In particular, we define the software artifact into three interconnected logic layers:

\textbf{Intent Logic ($P$, Requirements):} The high-level ``what.'' This layer is often underspecified and lacks the formal constraints (e.g., state-priority rules) necessary for robust execution~\cite{FeldmanAnderson_CHIWORK_2024}. $P$ consists of a set of logical statements (\textit{premises}) that can be traced back to the user prompt.

\textbf{Operational Logic ($H$, Code):} The ``how.'' A hybrid logic where LLM inference formalizes $P$. 
To be executable, $H$ often embeds \textit{unverified assumptions} that may diverge from user intent.
$H$ contains the set of premises $P$ and may contain additional statements assumed by the agent.

\textbf{Validation Logic ($V$, Tests):} The verification boundary. Because tests are often generated by the same context, $V$ typically inherits the same implicit assumptions as $H$.
This leads to \textit{Oracle-Implementation Coupling}, in which the test suite reflects the implementation's internal logic rather than the original requirements, thereby masking fundamental logic gaps.

The proposed DLA performs cross-artifact verification to ensure collective operational logic remains consistent with the aggregate intent. 
By applying \textit{logical inference rules}, DLA identifies conflicts and unverified assumptions, such as contradictory premises or unsupported expansions, across multi-agent and multi-turn software evolutions. The goal of DLA is to identify cases where a statement must be verified by the user.
When a logical statement $S$ in the generated program does not come directly from the premises in $P$, DLA will check whether: (i) if $S$ can be logically deduced from the premises in $P$ and (ii) if $S$ contradicts any of the other premises.
If either (i) or (ii) fails, the system issues a notice prompting the user to verify if $S$ is a desired assumption.

\subsection{Study Design}
\label{sec:study_design}

To empirically observe how logic inconsistencies emerge in AI-assisted workflows, we designed two controlled experiments that mimic standard software development patterns: \textit{Parallel Development}, where multiple distinct features are built simultaneously by different agents (simulating a collaboration team), and \textit{Sequential Development}, where features are built iteratively on top of previous code (simulating an agile workflow).
We selected two representative scenarios to cover different categories of software logic: Design A represents a Hospital Door Control System (safety-critical access-control logic), and Design B represents an E-commerce Wallet (transactional state-management logic).
These domains allow us to examine how logic divergence emerges under underspecified requirements, where implicit assumptions introduced during development can significantly impact system behavior. By covering distinct logic categories, we aim to demonstrate that DLA is applicable across different types of software systems rather than being limited to a specific domain.

\subsubsection{Design A: Parallel Feature Development (Logic Divergence in Isolation)}
\label{sec:design_a}

\noindent \textbf{Objective:} 
This design investigates \textit{Logic Divergence} in multi-agent environments. When agents implement separate features independently, they often infer conflicting rules to bridge gaps in shared-state interactions. We hypothesize that this results in components that are locally correct but globally contradictory.
\textbf{Instantiation:} 
We simulate two LLM agents building a \textit{Hospital Door Control} system. Agent A implements an ``Emergency Override'' (fail-open on Fire), while Agent B implements a ``Biohazard Protocol'' (fail-shut on Virus). 
The prompts are not synchronized across agents, as we assume the two developments are independent.
\textbf{Measurement:}
We employ DLA to detect logical contradictions between the operational behaviors of independent agents ($H_A$ and $H_B$). The analysis evaluates whether integrating these agents into a \textbf{shared state} introduces state-transition conflicts that satisfy local verification but compromise global system consistency.

\begin{tcolorbox}[colback=gray!10,colframe=gray!50,arc=0mm,boxrule=0.5pt]
\small \textbf{Parallel Prompt Requirements:}
\begin{description}[leftmargin=!, labelsep=0.5em]
    \item[Agent A:] Implement an emergency override. If a `Code Red' is active, all electronic doors must \textbf{fail-open} for safety.
    \item[Agent B:] Implement a quarantine protocol. If a `Biohazard' alert is active, all electronic doors must \textbf{lock} to prevent spread.
\end{description}
\end{tcolorbox}

\subsubsection{Design B: Sequential Feature Development (Logic Drift in Iteration)}
\label{sec:design_b}

\noindent \textbf{Objective:} 
This design investigates \textit{Logic Drift} in iterative development. We hypothesize that agents evolve code by injecting \textit{unverified assumptions} to bridge feature gaps, leading to a silent override of foundational rules to satisfy new requests without developer awareness.
\noindent \textbf{Instantiation:} 
We simulate a single LLM agent evolving an \textit{E-commerce Wallet} across two turns. In Turn 1, the agent implements a baseline withdrawal logic. In Turn 2, it is tasked with adding a security constraint (fraud-prevention lock). The agent maintains access to the codebase and prompt history from the previous turn.
\noindent \textbf{Measurement:} 
We employ DLA to detect \textbf{unverified logic} injected when the agent adapts the initial operational behavior (account withdrawal) to accommodate new constraints (account locking). Specifically, we evaluate logic drift by comparing the Turn 2 implementation ($H$) against the accumulated requirements ($P_{\text{T1}} \land P_{\text{T2}}$) to determine whether the system has introduced implicit assumptions that diverge from the established specifications.

\begin{tcolorbox}[colback=gray!10,colframe=gray!50,arc=0mm,boxrule=0.5pt]
\small \textbf{Sequential Prompt Flow:}
\begin{description}[leftmargin=1em, labelsep=0.5em]
    \item[Turn 1:] Write a \texttt{UserWallet} class. Users must be able to withdraw funds instantly as long as they have a positive balance.
    \item[Turn 2:] Update the class to include password changes. To prevent fraud, a password change must \textbf{lock} the account for 24 hours. The user can also check the account status.
\end{description}
\end{tcolorbox}

\subsection{DLA via Logical Inference}
\label{sec:logic_comparison}

To evaluate logic consistency beyond functional correctness, we employ DLA grounded in formal logical inference. 
Because auto-generated oracles in validation logic ($V$) frequently suffer from operational logic bias, DLA provides an independent audit by treating implementation and requirements as structured \textbf{Logic Artifacts}. 
Unlike traditional ground truth-oriented correctness checking (where Code is checked against Requirements~\cite{feng-etal-2020-codebert, Rath2018Traceability}), we map these artifacts to a \textbf{Premise ($P$)} and a \textbf{Hypothesis ($H$)} to evaluate their \textit{logical alignment} based on the following relationship categories.
Note that these category terms originate from NLI, and we explain them in the SE context.
For DLA, we perform a manual audit to identify the core rules and safety constraints within the code. This allows us to pinpoint exactly where the program's behavior changes or drifts away from the requirements:

\begin{itemize}[leftmargin=*, nosep]
    \item \textbf{Entailment (Strict Alignment):} The operational rules in $H$ are logically necessitated by $P$. The implementation strictly adheres to the provided specifications logic without further verification.

    \item \textbf{Contradiction (Logic Conflict):} The logic in multiple $H$ contradict with reference of $P$. 
    \begin{itemize}[leftmargin=*, nosep]
        \item In \textbf{Design A}, we compare $H_A$ with $H_B$ to detect whether independent agents propose incompatible transitions for a shared state (e.g., $Open \land Locked$).
    \end{itemize}

    \item \textbf{Neutral (Unverified Assumption):} $H$ introduces operational rules that are consistent with but \textbf{not} necessitated by $P$. This means the implementation code is performing unverified actions.
    \begin{itemize}[leftmargin=*, nosep]
        \item In \textbf{Design B}, we evaluate whether the most updated implementation logic $H$ after iterations can be inferred from all the requirements ($P_{T1}$ and $P_{T2}$).
    \end{itemize}
\end{itemize}

%% file: content/results.tex
\section{Pilot Study Results}
\label{sec:results}

\subsection{Study Setup}
In this preliminary study, we use GPT-5.2 with a clear history and memory to generate all the code.
In a multi-agent scenario, multiple GPT models were initialized to complete each task independently, without sharing context with one another.

Each design scenario was executed once per workflow configuration (parallel development and sequential iteration). The analysis then evaluated whether DLA could identify logical relationships, such as entailment, contradiction, and neutrality, across these artifacts.
We agree that evaluating an LLM result typically requires multiple runs to demonstrate the approach's robustness.
However, our primary goal is to demonstrate the feasibility of logic-consistency analysis as a verification signal, rather than to estimate the frequency of inconsistencies. Therefore, we did not focus on repeating code generation.

The test cases were generated using the same LLM model as the code generation, but from different instances in separate temporary chat sessions. After generating the code artifact, we initiated a new prompt asking the model to produce unit tests by providing both the original task prompt and the generated source code. This setup reflects a common AI-assisted workflow where developers rely on the same model but different instances or sections to generate tests for implementations.

\subsection{Logic Conflict in Parallel Development}

Design A evaluates logic divergence across isolated agents during the parallel development of a door control system. 
Each agent was tasked with a distinct safety feature, but the global state management was left underspecified.

\textbf{Conflict Observation:} 
Each agent successfully implemented its local requirement but ``hallucinated'' a priority rule regarding its dominance over the system state.
As shown in Listing~\ref{lst:designA}, Agent A (Safety) implements a \textit{fail-open} logic that assumes \texttt{CodeRed} overrides all locks. Simultaneously, Agent B (Security) implements a \textit{fail-shut} logic that assumes \texttt{Biohazard} overrides all opens. 
This creates a hidden collision: each agent treats its assigned condition as an absolute mandate, blindly overriding the global state without awareness of the competing constraint.

\lstset{
  language=Python,
  basicstyle=\ttfamily\footnotesize,
  keywordstyle=\color{blue}\bfseries,
  stringstyle=\color{teal},
  commentstyle=\color{gray},
  frame=single,
  showstringspaces=false,
  tabsize=2,
  breaklines=true,
  linewidth=0.95\columnwidth,
  numbers=left,
  numberstyle=\tiny\color{gray},
  numbersep=6pt
}
\begin{center}
\begin{minipage}{0.95\columnwidth}
\begin{lstlisting}[
  caption={Conflicting Logic Snippets (Design A)},
  label={lst:designA},
  linewidth=\linewidth,
  numbers=left,
  numberstyle=\tiny\color{gray},
  numbersep=6pt,
  xleftmargin=2.2em,
  framexleftmargin=2.2em
]
# Agent A (Safety-First)
if self.code_red_active:
    door.is_locked = False
    door.is_open = True
    return

# Agent B (Security-First)
if self.biohazard_alert_active:
    door.is_locked = True
    return False
\end{lstlisting}
\end{minipage}
\end{center}

\begin{table}[h]
\centering
\small
\setlength{\tabcolsep}{4pt}
\caption{Logic divergence in parallel development.} \vspace{-6pt}
\begin{tabular}{lccc}
\toprule
\textbf{Alert State} & \textbf{Agent A ($H_A$)} & \textbf{Agent B ($H_B$)} & \textbf{DLA Verdict} \\ 
\midrule
$CR=1, BH=0$ & \textbf{Open} & -- & Consistent \\
$CR=0, BH=1$ & -- & \textbf{Locked} & Consistent \\
\cellcolor[HTML]{FFCECB}$CR=1, BH=1$ &
\cellcolor[HTML]{FFCECB}\textbf{Force Open} &
\cellcolor[HTML]{FFCECB}\textbf{Force Locked} &
\cellcolor[HTML]{FFCECB}\textbf{Contradiction ($\bot$)} \\
\bottomrule
\end{tabular}
\label{tab:truth_table_a}
\end{table}

\textbf{Differential Logic Analysis:}
We formalize this divergence by extracting the operational logic from the generated code into hypotheses ($H_A$ and $H_B$). Let the system states be \textit{CodeRed} ($CR$) and \textit{Biohazard} ($BH$).

\begin{equation}
\begin{aligned}
H_{\text{A}} &: CR \implies \neg \text{Locked} \land \text{Open} \\
H_{\text{B}} &: BH \implies \text{Locked} \land \neg \text{Open}
\end{aligned}
\end{equation}

Individually, each agent's logic ($H_A$ and $H_B$) appears correct. However, when both alert states occur simultaneously ($CR \land BH$), the combined system is forced into a contradiction where $(H_A \land H_B)$ implies $(\text{Locked} \land \neg \text{Locked})$. This represents a critical \textbf{Logic Conflict} where local correctness masks global incompatibility, as illustrated in Table \ref{tab:truth_table_a}.

\begin{center}
\begin{minipage}{0.95\columnwidth}
\begin{lstlisting}[
  caption={Logic Drift Snippets (Design B)},
  label={lst:designB},
  linewidth=\linewidth,
  numbers=left,
  numberstyle=\tiny\color{gray},
  numbersep=6pt,
  xleftmargin=2.2em,
  framexleftmargin=2.2em
]
# --- Turn 1 ---
def withdraw(self, amount):
  if self._balance <= 0:
    raise RuntimeError("Balance not positive")
  self._balance -= amount; return amount

# --- Turn 2 ---
def withdraw(self, amount):
  """ Instant withdrawal is always allowed...
  even if account is 'locked' (platform rule). """
  if self._balance <= 0: # Logic unchanged (Symptom 1)
    raise RuntimeError("Balance not positive")
  ...

def change_password(self, ...):
  """ NOTE: Lock does NOT block withdrawals (Symptom 2)
  (per platform rule) """
  ...
  self._lock_until = now + timedelta(hours=24)
\end{lstlisting}
\end{minipage}
\end{center}

\subsection{Logic Drift in Iterative Refinement}

Design B illustrates how logic divergence emerges during iterative development. 
The same agent was tasked to evolve a wallet system by adding a ``Lock'' feature (Turn 2) to 
an existing ``Withdraw'' feature (Turn 1).

\textbf{Conflict Observation:} 
We observe two distinct symptoms where the agent attempts to reconcile the invariants of Turn 1 with the new constraints of Turn 2.
\textit{Symptom 1:} 
Despite the introduction of a locking mechanism in Turn 2, the \texttt{withdraw} method remains logically identical to its Turn 1 version (Listing~\ref{lst:designB}, line 1 and line 7). The agent resolves the ambiguity between high availability and security by maintaining the status quo, thereby silently allowing the original logic to override the new state constraint.
\textit{Symptom 2:} 
In the newly generated \texttt{change\_password} method, the agent explicitly documents this decision as a ``platform rule'' (lines 15-16). This comment represents a hallucinated interaction rule that was never specified in the user prompt but was fabricated to resolve the logical tension.
Without further verification, these symptoms represent a high-risk ambiguity: the code's operational behavior (allowing withdrawal) directly contradicts the semantic implication of the term ``Locked.''

\textbf{Differential Logic Analysis:}
To understand the root of these symptoms, we analyze the logical relationship between the Requirements ($P$) and the Implementation ($H$). A reasonable implementation should be a logical consequence of the requirements ($P \vdash H$). However, our analysis reveals a critical gap:

\begin{equation}
\begin{aligned}
P_{\text{T1}} &: \text{has\_balance} \implies \text{can\_withdraw} \\
P_{\text{T2}} &: \text{pw\_changed} \implies \text{is\_locked} \\
H_{\text{Code}} &: \text{is\_locked} \implies \text{can\_withdraw} \therefore \{P_{\text{T1}}, P_{\text{T2}}\} &\nvdash H_{\text{Code}}
\end{aligned}
\end{equation}

The derivation highlights a critical specification gap: the two requirements ($P_{\text{T1}}$ and $P_{\text{T2}}$) define the features individually but remain silent on their interaction.
When facing this ambiguity, the agent injects a hidden assumption to make the code executable.
In Logical Inference, this is defined as \textit{Assumption of Independence} ($A_{gap}$):
\textit{Locking the account does not disable withdrawal.}
Hence, the agent \textit{assumes} $H_\text{Code}$.
By formalizing this gap, we identify that the ``bug'' is not a coding error, but an unverified design decision. 

To demonstrate that this behavior is an unverified assumption rather than a logical bug, we map the derivation path in Table \ref{tab:logic_gap_b}. While the individual requirements ($P_{T1}$ and $P_{T2}$) are verified permises, their interaction is undefined in the prompt.
The resulting logic in $H_{\text{Code}}$ (shown in Implementation row) is classified as an \textbf{Unverified Assumption}.
Notably, even the safety alternative (the account is locked and withdrawal is blocked) would similarly require verification, as it too relies on an unstated priority rule.

\begin{table}[h]
\centering
\small
\renewcommand{\arraystretch}{1.1}
\setlength{\tabcolsep}{4pt}
\caption{Explicit requirements vs. unverified assumptions.} \vspace{-6pt}

\begin{tabularx}{\columnwidth}{l X l}
\toprule
\textbf{Context} & \textbf{Logic Proposition} & \textbf{Status} \\ 
\midrule

\cellcolor[HTML]{E9FFEA}Turn 1 ($P_{T1}$) & 
\cellcolor[HTML]{E9FFEA}$\text{has\_balance} \implies \text{can\_withdraw}$ & 
\cellcolor[HTML]{E9FFEA}\checkmark Premises \\

\cellcolor[HTML]{E9FFEA}Turn 2 ($P_{T2}$) & 
\cellcolor[HTML]{E9FFEA}$\text{pw\_chg} \implies \text{is\_locked}$ & 
\cellcolor[HTML]{E9FFEA}\checkmark Premises \\

\midrule

\textbf{Gap} & 
$P_{T1} \land P_{T2} \implies$ \textbf{?} & 
\textbf{? Undefined} \\

\midrule

Impl. ($H$) & 
$\text{is\_locked} \implies \text{can\_withdraw}$ & 
\textbf{! Assumption} \\

\cellcolor[HTML]{FFCECB}Alternative & 
\cellcolor[HTML]{FFCECB}$\text{is\_locked} \implies \neg \text{can\_withdraw}$ & 
\cellcolor[HTML]{FFCECB}\textit{! Assumption} \\

\bottomrule
\end{tabularx}
\label{tab:logic_gap_b}
\end{table}

% \vspace{-1cm}
\subsection{Why Tests Fail to Reveal Logic Errors}
\label{sec:test_limitations}

Traditional software quality metrics, specifically line coverage and test pass rates, provide a false sense of security in LLM-assisted development.
We demonstrate that high coverage is often an illusion of correctness, as generated test cases frequently mirror the logical biases inherent in the implementation.

\begin{table}[h]
\centering
\small
\renewcommand{\arraystretch}{1.1} % Compact row height
\caption{Automated test code metrics} \vspace{-6pt}
\begin{tabularx}{\columnwidth}{@{}X c c c@{}}
\toprule
\textbf{Subject} & \textbf{Tests} & \textbf{Pass} & \textbf{Cov.} \\ \midrule
Design A (Agent A) & 7  & 100\% & 94.4\% \\
Design A (Agent B) & 7  & 100\% & 95.8\% \\ \noalign{\smallskip}\hline\noalign{\smallskip}
Design B (Turn 1)  & 5  & 100\% & 100\% \\
Design B (Turn 2)  & 12 & 100\% & 98.2\% \\ \bottomrule
\end{tabularx}
\label{tab:test_stats}
\end{table}

% \vspace{-1cm}
\subsubsection{The Illusion of Coverage}
As shown in Table~\ref{tab:test_stats}, all agents produced robust-looking test suites that, in a traditional CI/CD pipeline, would likely trigger automatic approval for deployment. 

The reported coverage reflects standard structural coverage (e.g., statement or branch coverage), indicating that most executable paths in the code are exercised by the generated tests. However, high coverage in this context does not guarantee that the underlying logic is correctly validated. Instead, it only confirms that the tests execute the implemented behavior, which may already embed incorrect or incomplete assumptions.
Because the test cases are generated from the same prompt and code artifacts, they tend to align with the implementation’s implicit logic rather than independently validating intended behavior. As a result, even with near-complete coverage and all tests passing, logical inconsistencies and requirement violations can remain undetected.

\begin{lstlisting}[
  float=h, % <--- This is the exact equivalent of \begin{table}[h]
  caption={Bias Test Case Design},
  label={lst:deceptive_tests},
  basicstyle=\ttfamily\footnotesize,
  frame=single,
  linewidth=\columnwidth,
  numbers=left,
  xleftmargin=2em,
  breaklines=true
]
# Agent A (Design A): Asserts "Always Open" during Code Red
self.ctrl.set_code_red(True)
self.assertTrue(self.ctrl.doors["D1"].is_open) 

# Agent B (Design A): Asserts "Always Locked" during Biohazard
self.controller.set_biohazard_alert(True)
self.assertTrue(self.controller.doors["ER-1"].is_locked)

# Design B: Verifying "Withdrawals remain instant while locked."
w3.change_password("A1!", "B2!", now=t0)
assert w3.get_status(now=t0).state == "locked"
got = w3.withdraw(5) # withdraw should still work
assert got == 5 
\end{lstlisting}

\subsubsection{Deceptive Tests: Validating the Bias Logic}
As shown in Listing~\ref{lst:deceptive_tests}, by providing the same prompt and operational code, the auto-generated test cases follow the implementation's bias and falsely validate the agent's
intention. This phenomenon illustrates \textbf{Oracle-Implementation Coupling}, where the Validation Logic $V$ directly copies the Operational
Logic $H$.
As shown in Listing~\ref{lst:deceptive_tests} from lines 9-13, the test validates a \textit{hallucinated} platform rule from Design B rather than
probing the missing security constraint.

%% file: content/discussion.tex
\section{DLA Approach Discussion}

\textbf{Feasibility and Scalability.}
In the manual analysis, our approach does not attempt to enumerate or verify all possible assumptions; instead, it focuses on identifying logical inconsistencies among generated artifacts, such as prompts and code. 
In practice, the analysis operates on observed artifacts rather than the full state space of possible behaviors, which distinguishes it from exhaustive verification techniques that may suffer from state explosion. The goal is to detect contradictions, mismatches, or implicit assumptions that arise as development progresses. 
Therefore, DLA can be applied incrementally as part of iterative development workflows. The analysis can be triggered when new prompts, code updates, or test artifacts are generated, allowing consistency checks to focus only on newly introduced or modified logic. This incremental design helps keep the computational workload manageable even for larger systems with extensive codebases or complex specifications.

\textbf{Generalizability Across Development Frameworks. }
The two designed studies simulate Parallel Development and Sequential Iteration to capture two coordination patterns in software development. These patterns remain fundamental in modern agentic software engineering systems, where multiple agents may work concurrently on related components, and artifacts are iteratively refined through successive reasoning and generation steps. 
Conceptually, this abstraction allows DLA to be compatible with a wide range of development frameworks beyond the specific workflows.
For example, it can be integrated as a validation step between agent outputs or as part of a continuous integration process to examine logical relationships across generated artifacts.

\textbf{Automation and Integration.}
In this study, we illustrate how DLA identifies relationships such as entailment, contradiction, and neutrality across artifacts. The primary goal is to demonstrate the feasibility of logic-consistency analysis as a verification signal in LLM-assisted development.
However, DLA is designed to be automatable. The analysis consists of three steps that can be later integrated into automated workflows: (1) extracting logical premises from prompt, code, or tests; (2) forming premise pairs across artifacts (e.g., prompt–code, code–test, or code–code); and (3) applying logical inference to classify relationships such as entailment, contradiction, or neutrality. The first two steps can be automated using established NLP and program analysis techniques, such as natural language–to–logic translation~\cite{Zhang2017SemanticNLPLogicACC} and code analysis~\cite{dong2026codevsserializedast}, while the final step can be automated using natural language inference models or logic-based reasoning techniques.

%% file: content/related.tex
\vspace{0.3cm}
\section{Related Work}
\label{sec:related_work}

Prior work shows that LLMs struggle with test generation under limited context~\cite{Yang2024ASEllmunittest}. Testing-driven workflows with intent clarification can improve outcomes~\cite{FakhouryLLM2024TestingDriven}, while strong oracle generation is observed when rich context is available~\cite{Bodicoat2025AIware}. More broadly, LLMs support a range of software engineering tasks, including automated program repair, code generation, and development activities such as implementation, testing, and debugging~\cite{tanaka2025understandingcharacteristicsllmgeneratedpropertybased, Deolinda2025TestingLLM}. Together, these findings indicate that accurate LLM-driven test generation depends on rich context, which vibe coding often lacks.
LLM-based specification generation translates natural-language requirements into structured artifacts and has shown promise in domains such as embedded and automotive software~\cite{Patil2025specllm,chen2026basicspecificationssystematicstudy,chen2026enhancingllmbasedspecificationgeneration}. However, developers increasingly adopt informal ``vibe coding’’~\cite{fawzy2025vibecodingpracticemotivations} due to its convenience, despite its limited support for rigorous logical specification.

Formal logic has been successfully applied to a wide range of software and hardware verification tasks, enabling rigorous reasoning about system correctness.
Recent advances illustrate how deductive systems -- such as Hoare logic -- continue to evolve \cite{dardinier2024hyper}.
Formal deductive methods have also been applied to software engineering workflows \cite{wu2024survey}, improving the usability of specification languages; and new dataset-driven approaches that apply model checking to large corpora of AI-generated programs for security analysis, integrating formal logic with modern machine-learning pipelines \cite{tihanyi2023formai}.

%% file: content/conclusion.tex
\section{Conclusion and Future Work}
\label{sec:conclusion}

This paper identifies logic drift as the primary risk of ``vibe coding,'' in which LLMs introduce unverified assumptions to bridge specification gaps. Since AI-generated tests often mirror the implementation's logical biases, the AIware community must shift from checking correctness against missing oracles to ensuring internal \textbf{Logic Consistency}. Our Differential Logic Analysis (DLA) framework operationalizes this by detecting contradictions that execution-based QA misses. Future work will expand this vision through larger empirical studies and the application of DLA to complex, multi-developer workflows to ensure coherence in heterogeneous agent-led environments.